# A Generalized Stability Analysis Method with Dynamic Phasors for LV AC Microgrids

Bülent Dağ*

Gazi University, Department of Electrical-Electronics Engineering, Ankara, 06570, TURKEY

*Abstract*—Representation of inductive coupling lines with conventional static phasors is the main reason of inadequacy of the existing phasors based simplified stability analysis methods for microgrids with inductive coupling lines. In the literature, dynamic phasors have been proposed for the dynamic modelling of inductive lines to conserve the simplified structure of the analysis method. In this study a generalized stability analysis method for LV AC microgrids, composed of droop controlled inverters, is presented. The proposed analysis method is based on the inclusion of dynamic phasors for inductive coupling lines into the existing phasors based stability analysis method. The results show that the stability analysis method with dynamic phasors successfully predicts the instability boundaries of LV AC microgrids.

## I. INTRODUCTION

MICROGRID is a special distributed generation application that provides grid disconnected (islanded) operation as well, in a controlled manner. In islanded mode, droop based voltage and frequency controls are commonly applied methods to the grid interfacing inverters of master Distributed Generators (DGs), to provide reliable and low cost (communicationless) energy management capability to the microgrid [1-8]. On the other hand, secondary (or slave) units without specific power capacities, such as renewable energy sources, are operated in P-Q control mode to supply available power [1-3,8]. P-Q controlled units do not serve for the voltage and frequency regulation of the microgrids.

Small-signal stability of the microgrids has been one of the common research subject recently, as it enables selection of relevant system parameters maintaining steady operation under normal network conditions (i.e. under small disturbances) [9]. Due to more complicated control structures of DG inverters of microgrids compared to conventional generators, an accurate enough but also practical stability analysis method is required. In the literature, there exist a number of studies that constitute several small-signal stability analysis approaches for microgrids [10]-[12]. The one presented in [10] analyses stability of a stand-alone multi-inverter system by taking into account only power dynamics of ideal droop controlled DG inverters. It neglects inner voltage and current control dynamics of DG inverters and also inductive coupling dynamics of interconnecting lines. When properly tuned, comparatively fast voltage and current dynamics have little effect on system stability [13], however assumption of static coupling impedances reduces accuracy of the presented method for microgrids with inductive coupling lines [14], [15]. In addition, as in all referred studies, for the determination of equilibrium points of the small-signal model, additional static load flow analysis is required. This further increases the effort required for the analysis. A couple of detailed analyzes, taking into account also the voltage and current dynamics with dynamic inductive lines, are presented in [11] and [12]. Although the presented analysis methods are very enlightening for understanding the origin of different eigenvalues of the system and their effects on the system stability, they are quite far from the required generality and simplicity to be used for large microgrids. And also, as mentioned before, the requirement of the equilibrium points for the small-signal model increases the effort required for the analysis further. Due to the complexity of the detailed analysis methods, there exist attempts to reduce order of the large microgrid systems in the literature. As an example, the study presented in [16] neglects the power filter dynamics and coupling impedances (assuming comparatively large interconnecting line impedances) to simplify the analysis. However, power filters with slow dynamics may affect the system stability considerably and should not be neglected [10]. Besides, the large interconnecting line impedance assumption can not be applied to distribution networks where microgrids are applied more frequently. To overcome the reduced accuracy due to static coupling impedances, the study presented in [14] applies dynamic phasors in modelling inductive line dynamics. But the study assumes stiff AC grid with constant voltage so the model emerges as a single inverter connected to a stiff AC grid. Apparently, the proposed model in this manner is quite far from the actual model of the islanded microgrids with many DGs connected to a variable voltage grid. On the other hand, dynamic phasors can be a proper solution to simplify the modelling and stability analysis of the microgrids [17]. There exist several other approaches for model reduction of microgrid systems, based on singular perturbation method [18]-[20]. These methods have respective accuracies depending on the level of reduction, but the accuracy of the model decreases as the level of reduction increases. Moreover, all mentioned methods are dependent on the applied microgrid topology and so do not provide a generalized approach.

A generalized and simplified stability analysis method for LV microgrids that is based on conventional phasors has been presented in [15]. In this method, the inner voltage and current control dynamics are neglected, the power dynamics are modelled with conventional voltage and impedance phasors and for the generalization an approximately equivalent common microgrid structure is proposed that can be applied to any LV microgrid configuration with any size. This network simplification relies on the respective low interconnecting line impedances of the LV networks. In addition, identification of the equilibrium points of state variables process is eliminated by analytical methods. In this process, the equilibrium points are identified from static analysis of the microgrid, including droop equations, as well. As a result, a simplified and generalized stability analysis method for LV microgrids has been constructed. However, the main drawback of the proposed method is decrease in accuracy in case of dynamic (inductive) coupling impedances. As shown in [15], while the method has quite accurate results for resistive couplings, the accuracy reduces as the coupling impedances get inductive. This is due to the use of static model for coupling impedances in the method, as also mentioned in [10], [14]. A dynamic phasors method was applied in [14] to increase the accuracy, however as mentioned before, the presented study can be applied to a single DG connected to a stiff AC grid.

---

*e-mail: bulentdag@gazi.edu.tr

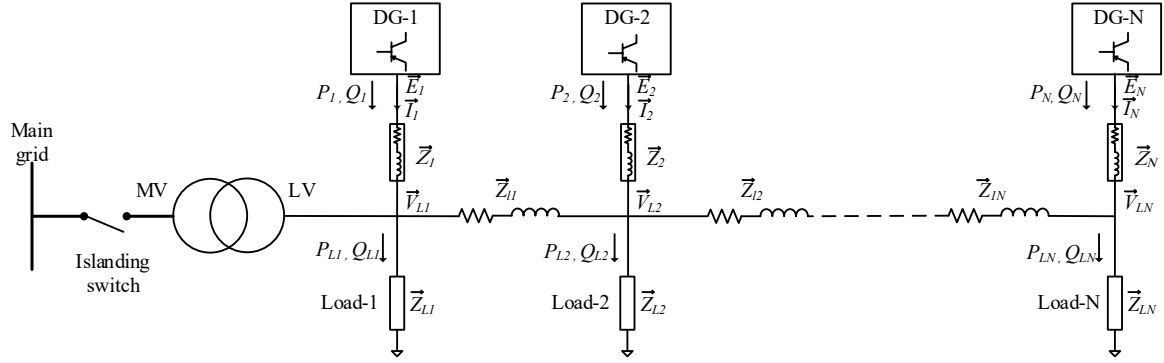

**Fig. 1** LV microgrid system

In this study, the dynamic phasors method is adopted to the simplified and generalized stability analysis method presented in [15]. Implementation of dynamic phasors leads current magnitudes and phases to emerge as new dynamic variables, providing more accurate representation of microgrids, while conserving the generalized and simple structure of the method presented in [15]. The results show that with the inclusion of dynamic phasors the proposed analysis method now can successfully detect the instability boundaries of microgrids with inductive couplings. The remainder of the paper is configured as follows; In Section II, microgrid and DG inverter models are described briefly. Generalized microgrid model with dynamic phasors is described in detail in Section III. Section IV gives the analysis results and comperes them with the simulation results. And finally, the conclusion section is given in Section V.

## II. MICROGRID AND DG INVERTER MODELS

### A. Generalized LV Microgrid Model

In Fig.1, a generic microgrid structure, for practical LV distribution network applications, is shown. Although it is in radial form, the method can be applied to networks in ring form, as well. The islanding switch is placed on the high voltage side of the distribution transformer for safety reasons [21]. The DGs are connected to the microgrid via coupling impedances for stability reasons to provide enough interconnection impedance between parallel operating DGs. They can either be grid side physical inductances of output LCL filters or virtually implemented impedances [22]. In Fig. 1, $\vec{E}_i$ and $\vec{I}_i$ represent complex output voltages and currents of DGs and $\vec{V}_{Li}$ correspond to grid side voltage of $i^{th}$ DG ($i = 1,2,...,N$). $\vec{Z}_i$ is the coupling impedance of the corresponding DG-i and $\vec{Z}_{li}$ is the impedance of the respective part of the main LV distribution feeder. $P_i$ and $Q_i$ correspond to the active and reactive output powers of DGs, respectively, while $P_{Li}$ and $Q_{Li}$ are those of distributed loads through the microgrid with corresponding load impedances of $Z_{Li}$. $N$ is the total number of active nodes in the microgrid where DGs are connected.

As described in [15], in LV implementation of microgrids the line impedances of distribution networks can be very small compared to the coupling impedances of DGs. And actually, in practice the coupling impedances should be large enough to provide enough stability margin for parallel operating DGs [11], [22], [23]. Therefore, based on this consideration, to simplify and generalize the microgrid network the interconnection line impedances of the distribution feeder can be neglected. The resulting network, emerging from this neglection, is shown in Fig. 2. In this simplified form, the coupling impedances, $Z_i = r_i + j\omega L_i$ are the same as those of the actual network and $V_L$ and $\emptyset_L$ are amplitude and phase of the unified load voltage. The unified load impedance here is the equivalent impedance of all loads in parallel in the microgrid, where,

$$\begin{cases} P_L = P_{L1} + P_{L2} + \cdots + P_{LN} \\ Q_L = Q_{L1} + Q_{L2} + \cdots + Q_{LN} \end{cases} \quad (1)$$

Note that by neglecting the distribution feeder line impedances, any LV microgrid configuration can be approximately represented with the equivalent form given in Fig. 2, to which, the stability analysis method can be applied now in a generalized manner. On the other hand, for HV networks where line impedances are significant and can not be neglected, approximate Kron Reduction methods can be applied to represent the networks in the form of Fig. 2 [24]. However, this issue is out of scope of this paper. In the following section the DG inverter model, based on droop controlled Voltage Source Inverter (VSI), will be described briefly.

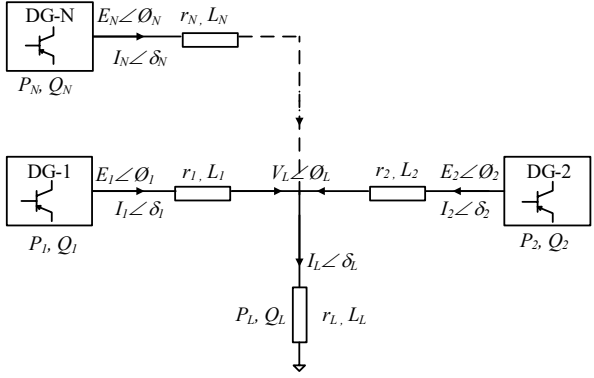

**Fig. 2** Simplified LV microgrid model

*B. DG Inverter Model*

Block diagram of a droop controlled DG inverter is shown in Fig. 3. The common approach for the inner voltage and current controllers is the use of PI controllers [11], [25], [26]. Their performance is quite well especially with the introduction of load compensating feed forward method presented in [11] for voltage controller. By this way, the inherent output impedance effect of the voltage controller can be eliminated and the effect of external coupling impedance on system performance can be directly analyzed. On the other hand, when properly tuned, voltage and current dynamics have not significant effect on overall system stability in terms of power dynamics since relevant system eigenvalues are located very far from critical instability region, as shown in the detailed model in [11]. Therefore, only the power dynamics dictated by droop control are concerned in this study, so an analysis in the complex domain with dynamic phasors can be sufficient.

Droop control takes place in Power Control Block in Fig. 3, where, frequency of the voltage sources is regulated with active power while amplitudes are regulated with reactive power in the form of well-known droop equations as below,

$$\omega = \omega_s - mP \quad (2)$$
$$E = E_s - nQ \quad (3)$$

Here, $\omega$ and $E$ are angular frequency and amplitude of the inverter reference output voltage $v_o^*$ (see Fig. 3) with set values of $\omega_s$ and $E_s$ corresponding to maximum values for frequency and voltage amplitude, respectively. $P$ and $Q$ are inverter's filtered active and reactive output powers and $m$ and $n$ are droop coefficients for frequency and amplitude, respectively. Power stage of a DG consists of a three-leg inverter, an *LC* filter and a coupling impedance *Z* at the output. Coupling impedances are required especially in LV applications for stability purposes because as shown in [11] and [15] stability of inverter-based microgrids requires coupling impedances above a critical value between the droop controlled VSIs. They can be implemented either with real inductive impedances or virtually by the use of a control implementation shown with dashed lines in Fig. 3 [22].

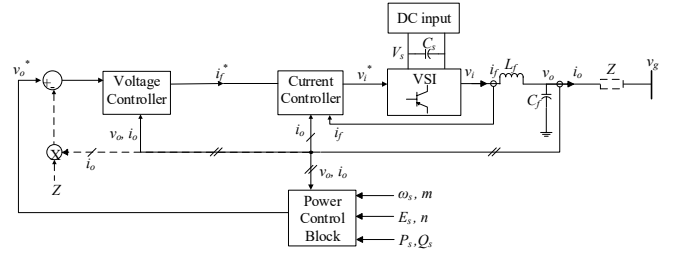

**Fig. 3** Electrical circuit and control block diagram of a droop controlled VSI.

### III. GENERALIZED STABILITY ANALYSIS WITH DYNAMIC PHASORS

In this section, equations of the proposed small-signal stability analysis method, based on the simplified microgrid model in Fig. 2, DG inverter model and dynamic coupling phasors, are derived. Since the system is non-linear the analysis is held in small-signal sense around corresponding steady-state operating conditions. The analytical methods for approximating equilibrium points are also described.

*A. DG Inverter Equations*

In the simplified microgrid model in Fig. 2 the active and reactive powers injected from the $i^{th}$ DG ($i = 1,2,..,N$) can be expressed in phasor form as follows,

$$P_i = pRe(\bar{V}_L \bar{I}_i^*) = pV_L I_i \cos(\emptyset_L - \delta_i) \quad (4)$$
$$Q_i = pIm(\bar{V}_L \bar{I}_i^*) = pV_L I_i \sin(\emptyset_L - \delta_i) \quad (5)$$

where, $p$ is 3/2 for 3-phase and 0.5 for single-phase networks. Note that the power expressions in (4) and (5) does not include coupling line losses. Linearized power expressions wrt. $V_L$, $\emptyset_L$, $I_i$ and $\delta_i$ around equilibrium values $V_{Le}$, $\emptyset_{Le}$, $I_{ie}$, $\delta_{ie}$ are obtained as follows,

$$\Delta P_i = k_{i1}\Delta V_L + k_{i2}\Delta\emptyset_L + k_{i3}\Delta I_i + k_{i4}\Delta\delta_i \quad (6)$$
$$\Delta Q_i = k_{i5}\Delta V_L + k_{i6}\Delta\emptyset_L + k_{i7}\Delta I_i + k_{i8}\Delta\delta_i \quad (7)$$

where,

$$\begin{cases} k_{i1} = pI_{ie}\cos(\emptyset_{Le} - \delta_{ie}) = \frac{P_{ie}}{V_{Le}} \\ k_{i2} = -pV_{Le}I_{ie}\sin(\emptyset_{Le} - \delta_{ie}) = -Q_{ie} \\ k_{i3} = pV_{Le}\cos(\emptyset_{Le} - \delta_{ie}) = \frac{P_{ie}}{I_{ie}} \\ k_{i4} = pV_{Le}I_{ie}\sin(\emptyset_{Le} - \delta_{ie}) = Q_{ie} \\ k_{i5} = pI_{ie}\sin(\emptyset_{Le} - \delta_{ie}) = \frac{Q_{ie}}{V_{Le}} \\ k_{i6} = pV_{Le}I_{ie}\cos(\emptyset_{Le} - \delta_{ie}) = P_{ie} \\ k_{i7} = pV_{Le}\sin(\emptyset_{Le} - \delta_{ie}) = \frac{Q_{ie}}{I_{ie}} \\ k_{i8} = -pV_{Le}I_{ie}\cos(\emptyset_{Le} - \delta_{ie}) = -P_{ie} \end{cases} \quad (8)$$

The equivalent expressions of coefficients $k_{i1\sim8}$ on the right side of (8) are based on the fact that the power expressions in (4) and (5) are valid also in steady-state with equilibrium values

($P_{ie}$ and $Q_{ie}$ for powers).

As a common approach the DG powers in (4) and (5) are low pass filtered firstly and then applied in the droop control given in (2) and (3). Then, from (2) and (3) with filtered DG powers, the linearized droop equations are obtained as follows,

$$\Delta\omega_i = -m_i \frac{w_{fi}}{s+w_{fi}} \Delta P_i \quad (9)$$

$$\Delta E_i = -n_i \frac{w_{fi}}{s+w_{fi}} \Delta Q_i \quad (10)$$

where, $w_{fi}$ is cut-off frequency of the power filters. From (6) and (7) in combination with (9) and (10), state equations of the droop controlled DG inverters are obtained as follows,

$$\begin{cases} \Delta\dot{\omega}_i = -w_{fi}\Delta\omega_i + k'_{i1}\Delta V_L + k'_{i2}\Delta\emptyset_L + k'_{i3}\Delta I_i + k'_{i4}\Delta\delta_i \\ \Delta\dot{E}_i = -w_{fi}\Delta E_i + k'_{i5}\Delta V_L + k'_{i6}\Delta\emptyset_L + k'_{i7}\Delta I_i + k'_{i8}\Delta\delta_i \\ \Delta\dot{\emptyset}_i = \Delta\omega_i \end{cases} \quad (11)$$

where,

$$\begin{cases} k'_{i1} = -m_i w_{fi} k_{i1}, \quad k'_{i2} = -m_i w_{fi} k_{i2} \\ k'_{i3} = -m_i w_{fi} k_{i3}, \quad k'_{i4} = -m_i w_{fi} k_{i4} \\ k'_{i5} = -n_i w_{fi} k_{i5}, \quad k'_{i6} = -n_i w_{fi} k_{i6} \\ k'_{i7} = -n_i w_{fi} k_{i7}, \quad k'_{i8} = -n_i w_{fi} k_{i8} \end{cases} \quad (12)$$

The last term in (11), emerges from $\emptyset_i = \int \omega_i dt$.

### B. Coupling Line Equations with Dynamic Phasors

A dynamic phasor, $X\angle\delta = Xe^{j\delta}$ is characterized by a variable amplitude $X$ in contrary to the constant amplitude of conventional phasors. Then, derivative of a dynamic phasor appears as follows,

$$\frac{d}{dt}X\angle\delta = (\dot{X} + jX\dot{\delta})\angle\delta \quad (13)$$

It should be noticed that the first term in the parenthesis in (13) does not exist conventionally. Then, according to (13), and from Fig. 2, the coupling line equations are obtained as follows,

$$E_i\angle\emptyset_i - V_L\angle\emptyset_L = (L_i\dot{I}_i + jL_iI_i\dot{\delta}_i + r_iI_i)\angle\delta_i \quad (14)$$

where, $L_i$ and $r_i$ are inductance and resistance of the coupling line, respectively, corresponding to the $i^{th}$ DG. When the expression in (14), which is in polar form, is written in rectangular form, the following coupling line equations are obtained,

$$E_i Cos\emptyset_i - V_L Cos\emptyset_L = L_i\dot{I}_i Cos\delta_i - L_iI_i\dot{\delta}_i Sin\delta_i + r_iI_i Cos\delta_i \quad (15)$$

$$E_i Sin\emptyset_i - V_L Sin\emptyset_L = L_i\dot{I}_i Sin\delta_i + L_iI_i\dot{\delta}_i Cos\delta_i + r_iI_i Sin\delta_i \quad (16)$$

Rearranging (15) and (16) using appropriate trigonometric identities, the non-linear coupling line dynamic equations are obtained as follows,

$$\dot{I}_i = \frac{E_i}{L_i}\cos(\emptyset_i - \delta_i) - \frac{V_L}{L_i}\cos(\emptyset_L - \delta_i) - \frac{r_i}{L_i}I_i \quad (17)$$

$$I_i\dot{\delta}_i = \frac{E_i}{L_i}\sin(\emptyset_i - \delta_i) - \frac{V_L}{L_i}\sin(\emptyset_L - \delta_i) \quad (18)$$

Linearizing (17) and (18),

$$\Delta\dot{I}_i = k_{i9}\Delta E_i + k_{i10}\Delta\emptyset_i + k_{i11}\Delta I_i + k_{i12}\Delta\delta_i + k_{i13}\Delta V_L + k_{i14}\Delta\emptyset_L \quad (19)$$

$$\Delta\dot{\delta}_i = k_{i15}\Delta E_i + k_{i16}\Delta\emptyset_i + k_{i17}\Delta I_i + k_{i18}\Delta\delta_i + k_{i19}\Delta V_L + k_{i20}\Delta\emptyset_L \quad (20)$$

where,

$$\begin{cases} k_{i9} = \frac{\cos(\emptyset_{ie}-\delta_{ie})}{L_i} = \frac{P_{ie}}{pL_iE_{ie}I_{ie}} \\ k_{i10} = -\frac{E_{ie}}{L_i}\sin(\emptyset_{ie}-\delta_{ie}) = -\frac{Q_{ie}}{pL_iI_{ie}} \\ k_{i11} = -\frac{r_i}{L_i} \\ k_{i12} = \frac{E_{ie}}{L_i}\sin(\emptyset_{ie}-\delta_{ie}) - \frac{V_{Le}}{L_i}\sin(\emptyset_{Le}-\delta_{ie}) = I_{ie}\omega_{ie} \\ k_{i13} = -\frac{\cos(\emptyset_{Le}-\delta_{ie})}{L_i} = -\frac{P_{ie}}{pL_iI_{ie}V_{Le}} \\ k_{i14} = \frac{V_{Le}}{L_i}\sin(\emptyset_{Le}-\delta_{ie}) = \frac{Q_{ie}}{pL_iI_{ie}} \\ k_{i15} = \frac{\sin(\emptyset_{ie}-\delta_{ie})}{L_iI_{ie}} = \frac{Q_{ie}}{pL_iE_{ie}I_{ie}^2} \\ k_{i16} = \frac{E_{ie}}{L_iI_{ie}}\cos(\emptyset_{ie}-\delta_{ie}) = \frac{P_{ie}}{pL_iI_{ie}^2} \\ k_{i17} = -\frac{\dot{\delta}_{ie}}{I_{ie}} = -\frac{\omega_{ie}}{I_{ie}} \\ k_{i18} = -\frac{E_{ie}}{L_iI_{ie}}\cos(\emptyset_{ie}-\delta_{ie}) + \frac{V_{Le}}{L_iI_{ie}}\cos(\emptyset_{Le}-\delta_{ie}) = -\frac{r_i}{L_i} \\ k_{i19} = -\frac{\sin(\emptyset_{Le}-\delta_{ie})}{L_iI_{ie}} = -\frac{Q_{ie}}{pL_iV_{Le}I_{ie}^2} \\ k_{i20} = -\frac{V_{Le}}{L_iI_{ie}}\cos(\emptyset_{Le}-\delta_{ie}) = -\frac{P_{ie}}{pL_iI_{ie}^2} \end{cases} \quad (21)$$

When deriving equivalent expressions on right hand side of (21), it should be noted that in steady-state (equilibrium condition) $\dot{I}_{ie} = 0$ and $\dot{\delta}_{ie} = \omega_{ie}$ (steady-state microgrid frequency). Also, it should be noted that if coupling line losses are ignored, the power expressions in (4) and (5) can be defined alternatively as follows,

$$P_i = pRe(\bar{E}_i\bar{I}_i^*) = pE_iI_i\cos(\emptyset_i - \delta_i) \quad (22)$$

$$Q_i = pIm(\bar{E}_i\bar{I}_i^*) = pE_iI_i\sin(\emptyset_i - \delta_i) \quad (23)$$

Then, using (17), (18), (22) and (23) (or alternatively (4) and (5)), the coefficients can be equivalently derived in terms of known operating values and grid parameters as in (21). Therefore, as seen, all system coefficients, $k_{ij}$ ($i : 1…N, j : 1…20$) can now be obtained by simple analytical calculations using grid and DG parameters. This eliminates the additional grid simulation to determine the equilibrium points of the small-signal model and simplifies the analysis.

## C. Load Equations

Load equations relate the non-state variables $\Delta V_L$ and $\Delta \varnothing_L$ to the state variables $\Delta \omega_i$, $\Delta E_i$, $\Delta \varnothing_i$, $\Delta I_i$ and $\Delta \delta_i$. It should be noted that, as shown in [15], dynamic modelling of load inductance does not have a significant effect on the accuracy of the analysis, in contrary to the coupling impedances. Therefore, static modelling will be applied to obtain load equations as in [15]. According to Fig. 2, the total active and reactive load powers can be expressed as follows,

$$\begin{cases} P_L = \frac{V_L^2}{Z_L} cos\theta_L \\ Q_L = \frac{V_L^2}{Z_L} sin\theta_L \end{cases} \quad (24)$$

where, $Z_L \angle \theta_L$ is the unified load impedance. Linearizing (24),

$$\begin{cases} \Delta P_L = \frac{2V_{Le}}{Z_L} cos\theta_L \Delta V_L = \frac{2P_{Le}}{V_{Le}} \Delta V_L \\ \Delta Q_L = \frac{2V_{Le}}{Z_L} sin\theta_L \Delta V_L = \frac{2Q_{Le}}{V_{Le}} \Delta V_L \end{cases} \quad (25)$$

Neglecting line losses, the load powers can be expressed in terms of DG inverter powers as follows,

$$\begin{cases} \Delta P_L = \sum \Delta P_i \\ \Delta Q_L = \sum \Delta Q_i \end{cases} \quad (26)$$

Combining (6), (7), (25) and (26),

$$\frac{2P_{Le}}{V_{Le}} \Delta V_L = \sum k_{i1} \Delta V_L + \sum k_{i2} \Delta \varnothing_L + \sum k_{i3} \Delta I_i + \sum k_{i4} \Delta \delta_i \quad (27)$$

$$\frac{2Q_{Le}}{V_{Le}} \Delta V_L = \sum k_{i5} \Delta V_L + \sum k_{i6} \Delta \varnothing_L + \sum k_{i7} \Delta I_i + \sum k_{i8} \Delta \delta_i \quad (28)$$

where, all summations are from $i = 1$ to $N$. Eqns. (27) and (28) can be combined in matrix form as follows,

$$\begin{bmatrix} \Delta V_L \\ \Delta \varnothing_L \end{bmatrix} = \mathbf{A_L} \begin{bmatrix} \Delta \omega_i \\ \Delta E_i \\ \Delta \varnothing_i \\ \Delta I_i \\ \Delta \delta_i \end{bmatrix}_{5N \times 1} \quad (29)$$

where,

$$\mathbf{A_L} = \begin{bmatrix} \left(\frac{2P_{Le}}{V_{Le}} - \sum k_{i1}\right) & -\sum k_{i2} \\ \left(\frac{2Q_{Le}}{V_{Le}} - \sum k_{i5}\right) & -\sum k_{i6} \end{bmatrix}^{-1} \begin{bmatrix} [0] & [0] & [0] & [k_{i3}] & [k_{i4}] \\ [0] & [0] & [0] & [k_{i7}] & [k_{i8}] \end{bmatrix}_{2 \times 5N} \quad (30)$$

In (30), the submatrices $[k_{ij}]$ and $[0]$ are row vectors, defined as,

$$[k_{ij}]_{1 \times N} = [k_{1j} \quad \cdots \quad k_{Nj}] \text{ for } j = 3, 4, 7, 8 \quad (31)$$

$$[0]_{1 \times N} = [0 \quad \cdots \quad 0] \quad (32)$$

## D. Combined Microgrid Equations

Each DG inverter together with coupling impedances is a 5th order system with state variables $\Delta \omega_i$, $\Delta E_i$, $\Delta \varnothing_i$, $\Delta I_i$, $\Delta \delta_i$ resulting in a microgrid system with the order of $5N$. Combining (11), (19) and (20) with the substitution of (29), the resulting state-space representation of the microgrid system is obtained as follows,

$$\begin{bmatrix} [\Delta \dot\omega_i] \\ [\Delta \dot E_i] \\ [\Delta \dot\varnothing_i] \\ [\Delta \dot I_i] \\ [\Delta \dot\delta_i] \end{bmatrix}_{5N \times 1} = \mathbf{A_{sys}} \begin{bmatrix} [\Delta \omega_i] \\ [\Delta E_i] \\ [\Delta \varnothing_i] \\ [\Delta I_i] \\ [\Delta \delta_i] \end{bmatrix}_{5N \times 1} \quad (33)$$

where, system matrix $\mathbf{A_{sys}}$ is defined as follows,

$$\mathbf{A_{sys}} = \begin{bmatrix} [-w_{fi}] & [0] & [0] & [k'_{i3}] & [k'_{i4}] \\ [0] & [-w_{fi}] & [0] & [k'_{i7}] & [k'_{i8}] \\ [I] & [0] & [0] & [0] & [0] \\ [0] & [k_{i9}] & [k_{i10}] & [k_{i11}] & [k_{i12}] \\ [0] & [k_{i15}] & [k_{i6}] & [k_{i17}] & [k_{i18}] \end{bmatrix}_{5N \times 5N} + \begin{bmatrix} [k'_{i1}] & [k'_{i2}] \\ [k'_{i5}] & [k'_{i6}] \\ [0] & [0] \\ [k_{i13}] & [k_{i14}] \\ [k_{i19}] & [k_{i20}] \end{bmatrix}_{5N \times 2} \times \mathbf{A_L} \quad (34)$$

In (34), the submatrices of $5N \times 5N$ matrix are all diagonal matrices with the order of $N \times N$, and $[I]$ and $[0]$ are identity and null matrices, respectively. On the other hand, the submatrices of $5N \times 2$ matrix are all column vectors. In all of the submatrices $i$ ranges from 1 to $N$. The equilibrium points, required for the analysis, can be obtained approximately by using droop equations with relevant assumptions as described in [15]. The equilibrium points of the line currents can be obtained from the following equation,

$$I_{ie} = \frac{\sqrt{P_{ie}^2 + Q_{ie}^2}}{pE_{ie}} \quad (35)$$

## IV. ANALYSIS AND SIMULATION RESULTS

Developed stability analysis method has been applied to a test microgrid shown in Fig. 4. The test microgrid consists of three equivalent DGs with dedicated coupling impedances and three loads spread through a LV distribution feeder. Microgrid is a balanced three-phase network operating at 60 Hz with 180 Vpeak phase-neutral voltage. Parameters of the inverters and microgrid are given in Table 1. The three-leg, three-wire droop controlled VSIs of DGs have inner voltage and current controllers in d-q synchronous reference frame with feed-forward compensation (see Fig. 3). The power line characteristics of the test microgrid has been chosen the same as the main distribution feeder of CIGRE distribution test network which has a main feeder line impedance value of 0.163 +j0.136 Ω/km [21]. Total feeder length was kept as 500 m., which is a typical value for LV distribution feeders. For the simulations, which have been performed using Matlab-Simulink, the actual test microgrid in Fig. 4 has been used while the analysis has been applied to the simplified common microgrid model of the test microgrid by converting it to the form in Fig. 2, where the power line impedances are neglected. For the analysis, system state matrix $\mathbf{A_{sys}}$ in (34) is used from which eigenvalues ($\lambda_k$, $k = 1,..,5N$) of the system can be obtained. The eigenvalues can also be used to analytically observe behavior of the state variables by using the equation below, where $X_{ke}$ expresses equilibrium point and $X_k(0)$ expresses initial condition for the corresponding state variable, $X_k$. Initial conditions for $I_i$, $Ø_i$ and $\delta_i$ are zero and their equilibrium values can be obtained from (35), (22) and (4), respectively.

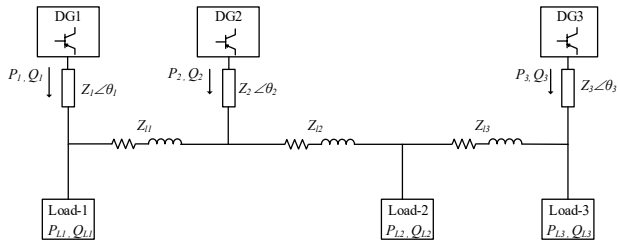

**Fig. 4** Structure of the test microgrid.

$$X_k = X_{ke} + \sum_{k=1}^{3N} C_k e^{\lambda_k t}, \quad C_k \ from \ given \ X_k(0) \quad (36)$$

In the first case, droop coefficients have been set small enough, as $m_1 = 2m_2 = 3m_3 = 2.5$ e-3, and $n_1 = 2n_2 = 3n_3 = 5$ e-3, so that the system operates in a stable region with enough margin. For comparison, stability analysis results with static phasors, presented in [15], are also included in the results. For the given condition above, analysis and simulation results of the DG inverter frequencies are shown in Fig. 5. The results show that both dynamic and static phasors based analysis methods predicts the system behavior in the stable regions, well enough.

To check the accuracy of the analysis methods around the instability regions, the analysis has been performed for the droop coefficients of $m_1 = 2m_2 = 3m_3 = 8.5$ e-3, and $n_1 = 2n_2 = 3n_3 = 5$ e-3. The analysis and simulation results of DG frequency variations are shown in Fig. 6. As seen, while the analysis method with static phasors has a stable behavior and so fails in predicting the instability boundary, the analysis with dynamic phasors predicts the instability boundary quite well. The frequency and amplitude differences of oscillations are due to the approximations made in the analysis, however, as seen from the results, dynamic phasors based analysis predicts the instability boundary quite well, with a response quite similar to simulation results. It should be noted that the stability analysis method has been developed in a generalized manner using Matlab m-files and can be adopted to different and very large microgrids easily.

## V. CONCLUSION

In this paper, a dynamic phasors based modelling has been applied for the stability analysis of the inverter based microgrids. A previous study which was based on conventional phasors was unsuccessful for the detection of instability boundaries of microgrid systems with inductive coupling lines. In the literature, dynamic phasors have been presented as a promising approach to increase accuracy of the static phasors based modelling and stability analysis methods while conserving the simple structure of the analysis method. The results showed that the dynamic phasors can successfully detect the instability boundaries of microgrids with inductive couplings. In addition, the analysis method has been presented in a generalized manner where a generalized microgrid structure is proposed for the analysis as in previous study. The generalized microgrid structure can be assumed accurate for LV distribution grids where interconnecting lines have negligible impedances compared to coupling impedances of microgrids. On the other hand, for the higher voltage networks where interconnecting lines can not be ignored, approximate Kron reduction techniques can be applied to convert the microgrid into the proposed generalized form. Finally, as in the previous study, the coefficients of the small-signal model developed are based on the calculable operating parameters of the microgrid system. This further simplifies the stability analysis procedure by eliminating the relevant simulations and analyses required to obtain equilibrium point values.

TABLE I
PARAMETERS OF TEST MICROGRID

| Parameter | Definition | Value (unit) |
|---|---|---|
| **DG Inverters and Coupling impedances** | | |
| $f_s$ | Switching frequency | 10 (kHz) |
| $L_f$, $r_f$ | Filter inductance and resistance | 1.84 (mH), 0.11 (Ω) |
| $C_f$ | Filter capacitance | 30 (uF) |
| $w_f$ | Power filters cut-off freq. | 31.85 (rad/sec) |
| $\omega_s$ | Frequency set value | 380 (rad/sec) |
| $V_{dc}$ | DC input voltage | 600 V |
| $L_1$, $r_1$ | Inductance and resistance of $Z_1$ | 1.57 (mH), 0.19 (Ω) |
| $L_2$, $r_2$ | Inductance and resistance of $Z_2$ | 2.46 (mH), 0.29 (Ω) |
| $L_3$, $r_3$ | Inductance and resistance of $Z_3$ | 2 (mH), 0.24 (Ω) |
| **Interconnection Lines** | | |
| $Z_{l1}$ | Impedance of 250 m line | 40.8 + j34 (m Ω) |
| $Z_{l2}$ | Impedance of 100 m line | 16.3 + j13.6 (m Ω) |
| $Z_{l3}$ | Impedance of 150 m line | 24.5 + j20.4 (m Ω) |
| **Loads (for 180 Vpeak grid voltage)** | | |
| $P_{L1},Q_{L1}$ | Active-reactive powers of Load 1 | 6 kW, 4 kVar |
| $P_{L2},Q_{L2}$ | Active-reactive powers of Load 2 | 2 kW, 1 kVar |
| $P_{L3},Q_{L3}$ | Active-reactive powers of Load 3 | 4 kW, 3 kVar |

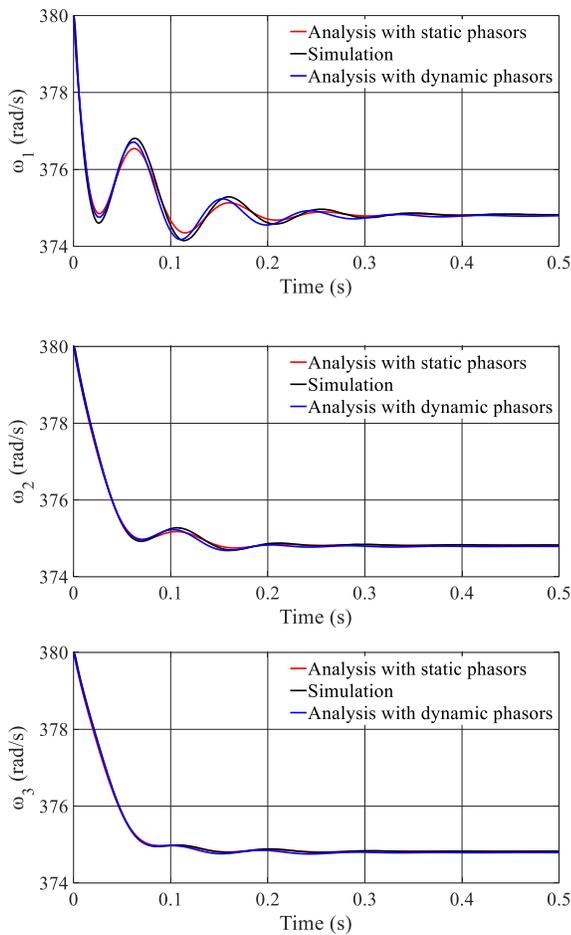

**Fig. 5** DG inverter frequencies for $m_1 = 2m_2 = 3m_3 = 2.5$ e-3, and $n_1 = 2n_2 = 3n_3 = 5$ e-3.

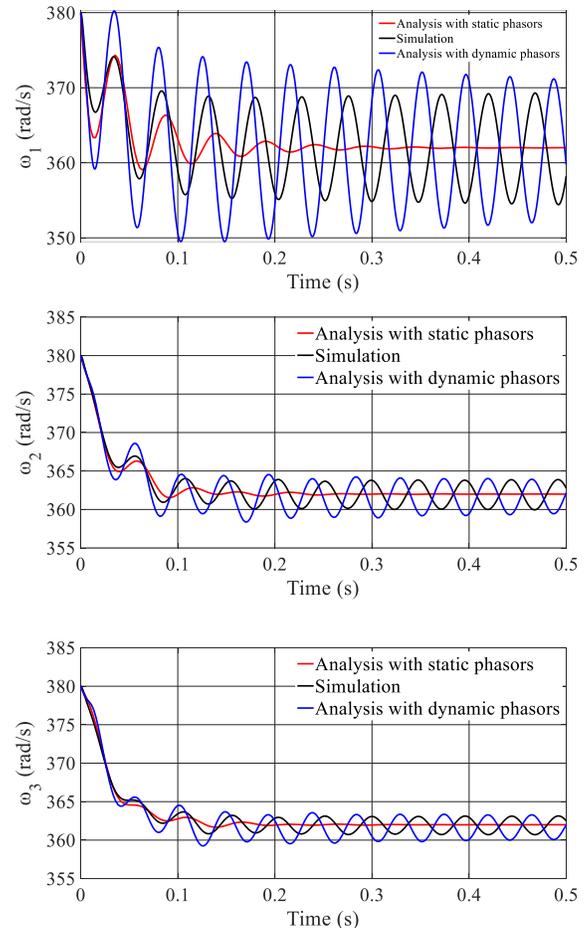

**Fig. 6** DG1 inverter frequency for $m_1 = 2m_2 = 3m_3 = 8.5$ e-3, and $n_1 = 2n_2 = 3n_3 = 5$ e-3.